\documentclass[preprint, 12pt]{elsarticle}
\usepackage{graphicx}
\usepackage{dcolumn}
\usepackage{amssymb}
\usepackage{bm}
\usepackage{enumerate}
\def\be{\begin{enumerate}}
\def\ee{\end{enumerate}}
\def\beq{\begin{equation}}
\def\eeq{\end{equation}}
\def\bea{\begin{eqnarray}}
\def\eea{\end{eqnarray}}

\def\g{\gamma}

\def\g{\gamma}

\def\3halfs{\textstyle{\frac{3}{2}}} 
\newcommand{\sla}[1]%
        {{\raise.15ex\hbox{$/$}\kern-.57em #1}}
\newcommand{\Sla}[1]%
        {{\raise.15ex\hbox{$/$}\kern-.75em #1}}

\def\ben{\begin{enumerate}}
\def\een{\end{enumerate}}
\def\bitem{\begin{itemize}}
\def\eitem{\end{itemize}}

\def\g{\gamma}

\journal{Astroparticle Physics}
\begin{document}
\begin{frontmatter}
\def\Universita{Universit\`a}
\title{A New Limit on Planck Scale Lorentz Violation from Gamma Ray Burst Polarization}
\author{Floyd W. Stecker}
\address{Astrophysics Science Division, NASA Goddard Space Flight Center, Greenbelt, MD 20771, USA}
\ead{floyd.w.stecker@nasa.gov}

\vspace{-1.5cm}
\begin{abstract}
Constraints on possible Lorentz invariance violation (LIV) to first order
in $E/M_{\rm Planck}$ for photons in the framework of
effective field theory (EFT) are discussed, taking cosmological factors
into account. Then, using the reported detection 
of polarized soft $\gamma$-ray emission from the $\gamma$-ray burst
GRB041219a that is indicative of an absence of vacuum birefringence, 
together with a very recent improved method for 
estimating the redshift of the burst, we derive  
constraints on the dimension 5 Lorentz violating modification 
to the Lagrangian of an effective local QFT for QED. Our new constraints 
are more than five orders of magnitude
better than recent constraints from observations of the Crab Nebula. 
We obtain the upper limit on the Lorentz violating dimension 5 EFT parameter
$|\xi|$ of $2.4 \times 10^{-15}$, corresponding to a constraint on the 
dimension 5 standard model extension parameter, 
$k^{(5)}_{(V)00} \le 4.2 \times 10^{-34}$ GeV$^{-1}$.
\end{abstract}
\end{frontmatter}
\newpage
\section{Introduction}

Because of the  problems associated with
merging relativity  with quantum  theory, it has  long been  felt that
relativity will have to be modified  in some way in order to construct
a quantum theory of gravitation. Since the Lorentz group is unbounded
at the  high boost (or high energy)  end, in principle it  may be
subject to  modifications in  the high boost limit ~\cite{CG,SG}. 
There is  also a
fundamental relationship between  the Lorentz transformation group and
the  assumption  that space-time  is  scale-free,  since  there is  no
fundamental length  scale associated with the  Lorentz group. However,
as noted  by Planck ~\cite{pl99},  there is a  potentially fundamental
scale associated  with gravity, {\it  viz.}, the Planck scale
$\lambda_{Pl}  = \sqrt{G\hbar /c^3} \sim 10^{-35}$ m,
corresponding  to  an  energy  (mass)   scale  of  $M_{Pl}  =  \hbar  c/
\lambda_{Pl} \sim 10^{19}$ GeV.

In recent years, there has been much interest in testing Lorentz invariance
violating terms that are of first order in $E/M_{Pl}$, since such terms
vanish at very low energy and are amenable to testing at higher energies.
In particular, tests using high energy astrophysics data have proved useful
in providing constraints on Lorentz invariance violation (LIV)
(e.g., see reviews in Refs.~\cite{jlm} and ~\cite{st09}).

\section{Vacuum Birefringence}

Important fundamental constraints on LIV come from searches for
the vacuum birefringence effect predicted within the framework
of the effective field theory (EFT) analysis of~\cite{MP}. (See also Ref.~\cite{CK}).
Within this framework, applying the Bianchi identities to the leading order Maxwell
equations {\it in vacua}, a mass dimension 5 operator term is derived of the form
\begin{equation}
{\Delta\cal{L}}_{\gamma} = {{\xi}\over{M_{Pl}}}{n^aF_{ad}n\cdot \partial (n_b \tilde{F}^{bd})}.
\label{deltaL}
\end{equation}
It is shown in Ref.~\cite{MP} that the expression given in Equation (\ref{deltaL}) is the only dimension 5 modification of the free photon Lagrangian that preserves both rotational symmetry and gauge invariance. This leads to a modification in the dispersion relation
proportional to $\xi (\omega/M_{Pl})=\xi (E/M_{Pl})$
\footnote{adopting the conventions $\hbar=1$ and the low
energy speed of light $c=1$.} with the new dispersion relation given by 
\beq
\omega^2 ~=~ k^2 \pm \xi\,  k^3/M_{Pl}. 
\label{disp-ph}\\
\eeq
Some models of quantized space-time suggest $\xi$ should be $\cal{O}$(1), (see, e.g., Ref.~\cite{ell}).
The sign in the photon dispersion relation
corresponds to the helicity,
i.e., right or left circular polarization. Equation~(\ref{disp-ph})
indicates that photons of opposite circular polarization
have different phase velocities and therefore travel with different speeds. 
The effect on photons from
a distant linearly polarized source can be constructed by
decomposing the linear polarization into left and right
circularly polarized states. It is then apparent that
this leads to a  rotation of the linear polarization direction
through an angle
\beq \theta(t)=\left[\omega_+(k)-\omega_-(k)\right]t_{P}/2~\simeq~\xi k^2 t_{P}/2M_{Pl}
\label{rotation}
\eeq
for a plane wave with wave-vector $k$, where $\xi k/M_{Pl} \ll 1$ and where
$t_{P}$ is the propagation time. 

Observations of
polarized radiation from distant sources can thus be
used to place an upper bound on $\xi$. 
The vacuum birefringence constraint arises from the fact that
if the angle of polarization rotation (\ref{rotation})
were to differ by more than $\pi/2$ over the energy range
covered by the observation
the instantaneous polarization
at the detector would fluctuate sufficiently
for the net
polarization of the signal to be suppressed well
below any observed value. 
The difference in rotation angles for wave-vectors $k_1$
and $k_2$ is
\beq
 \Delta\theta=\xi (k_2^2-k_1^2) L_{P}/2M_{Pl},
  \label{diffrotation}
\eeq
where we have replaced the propagation time $t_{P}$ by the propagation distance $L_{P}$
from the source to the detector. 

\begin{figure}

{\includegraphics[scale=0.9]{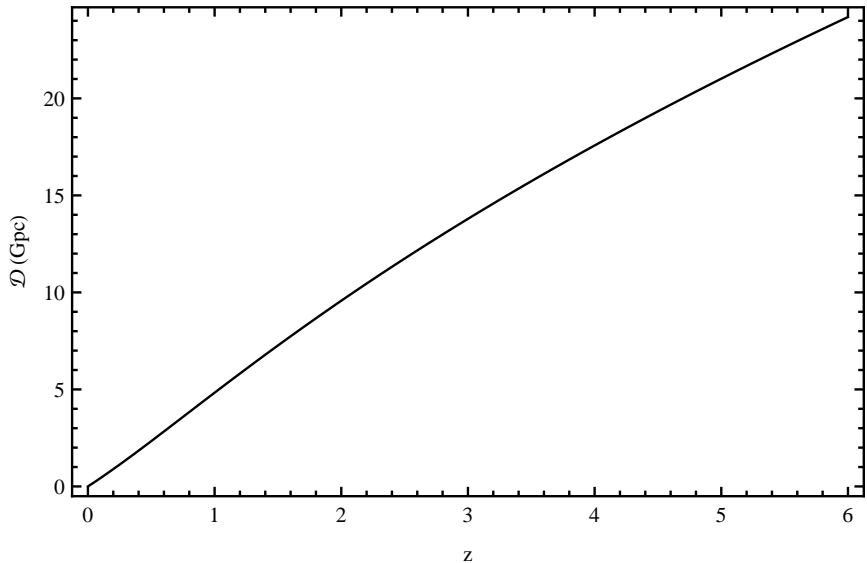}}
\caption{A linear plot of the integral $\cal{D}$ as defined in Equation (\ref{integral}), given as a function of redshift, $z$.}
\label{bigD}
\end{figure}

If polarization is detected from a source at redshift $z$, this yields the constraint 
\beq
   |\xi|< {{\pi M_{Pl}}\over{\int\displaylimits_{0}^{z} dz'{[k_2(z')^2-k_1(z')^2]|dL_{P}(z')/dz'|}}}
\label{constraint}
\eeq
where $k_{1,2}(z') = (1+z')\cdot k_{1,2}(z'=0)$
and
\beq
\Bigl{|}{dL_{P}\over{dz'}}\Bigr{|} = {{c}\over{H_{0}}}{{1}\over{(1+z')\sqrt{\Omega_\Lambda + (1+z')^3\Omega_m}}}.
\eeq
Defining
\beq
{\cal{D}} = {{c}\over{H_{0}}}\int\displaylimits_{0}^{z} dz' {{(1+z')}\over{\sqrt{\Omega_\Lambda + (1+z')^3\Omega_m}}}
\label{integral}
\eeq
it follows from equations (\ref{constraint})-(\ref{integral}) and the definitions of $k_{1,2}(z')$
that
\beq
|\xi|< {{\pi M_{Pl}}\over{{\cal{D}}(k_2^2-k_{1}^2)}},
\label{final}
\eeq
with the standard cosmological values~\cite{WMAP} of $\Omega_m = 0.27$, $\Omega_\Lambda = 0.73$, and $H_0$ = 71 km s$^{-1}$ Mpc$^{-1}$ (1 Mpc = $3.09 \times 10^{22}$ m). Figure~\ref{bigD} shows the function ${\cal{D}}(z)$ as defined in Equation (\ref{integral}).

\vspace{0.4cm}

\section{Previous Constraints}

A previous bound of $|\xi|\lesssim 2\times 10^{-4}$,
was obtained by Gleiser and Kozameh~\cite{GK} using the observed 10\%
polarization of ultraviolet light
from a  galaxy at distance of around 300 Mpc.
Fan et al. used the observation of polarized UV and optical radiation at several wavelengths from      
the $\gamma$-ray bursts (GRBs) GRB020813 at a redshift $z = 1.3$ and  
GRB021004 $z = 2.3$ to get a 
constraint of  $|\xi|\lesssim 2\times 10^{-7}$~\cite{FWX}.
Jacobson et al.~\cite{jlms} used a report of polarized $\gamma$-rays 
observed~\cite{GRBpol} in the
prompt emission from the $\g$-ray burst GRB021206 in the energy range 0.15 to 2 MeV 
using the {\it RHESSI} detector~\cite{rhessi} to place strong limits on $\xi$.
However, this claimed polarization detection has been refuted~\cite{RF,Wi}.
 
Kosteleck\'{y} and Mewes~\cite{KM09} have shown that the EFT model parameter $\xi$ can be
related to the model independent isotropic dimension 5 standard model extension (SME) parameter
$k^{(5)}_{(V)00}$. They derive the relation
\begin{equation}
k^{(5)}_{(V)00} = 3\sqrt{4\pi}\xi/5M_{Pl},
\label{km}
\end{equation}
which we use in this paper. Their upper limit of $1 \times 10^{-32}$ GeV$^{-1}$, obtained by assuming a lower limit on the redshift of these bursts of $z = 0.1$, 
then corresponds to the constraint
$\xi < 6 \times 10^{-14}$.\footnote{Ref.~\cite{kr11} gives a table of similar limits on $k^{(5)}_{(V)00}$ with citations.} 

More recently, Maccione et al. have derived a constraint of
$|\xi|\lesssim 9\times 10^{-10}$ using 
observations of polarized hard X-rays from the Crab Nebula
detected by the {\it INTEGRAL} satellite~\cite{Mac}. 

It is clear from Equation~(\ref{constraint}) that the
larger the distance of the polarized source, and the
larger the energy of the photons from the source, the
greater the sensitivity to small values of $\xi$.
In that respect, the ideal source to study would be
polarized X-rays or $\gamma$-rays from a GRB with {\it a
known redshift} at a deep cosmological distance~\cite{jlms}.

\section{A New Treatment}

Unfortunately, despite the
many GRBs that have been detected and have known host galaxy spectral redshifts,
none of these bursts have measured $\gamma$-ray polarization. However, in this 
paper we take a new approach, deriving an estimated redshift for GRB041219a. This is
a GRB with reported polarization but no spectral redshift measurement. 

Polarization at a level of 63(+31,-30)\% to 96(+39,-40)\% in the soft $\gamma$-ray energy range
has been detected by analyzing data from the spectrometer on {\it INTEGRAL}
for GRB041219a in the 100 to 350 keV energy range~\cite{Mc}. It should be
noted that that a systematic effect that might mimic
polarization in the analysis could not definitively be excluded. 
This GRB does not have an associated host galaxy spectral redshift.

Useful relations have been recently obtained where known spectral redshifts
of GRBs are statistically correlated with various observational parameters of the bursts
such as luminosity, the Band function~\cite{band} parameter $E_{peak}$, rise time, 
lag time and variability of a burst ~(Ref. \cite{wang} and references therein). 
A detailed treatment of these
correlations is given in Ref.~\cite{wang}. By deriving updated luminosity correlations for
a very large number of GRBs, they find the tightest correlation is the luminosity-$E_{peak}$
correlation. Using the relation given in Ref.~\cite{wang}, 
\beq
\log L = 51.75 + 1.35 \log [(1+z)E_{peak}/300 {\rm keV}]
\label{luminosity}
\eeq
and the iterative method described in Ref.~\cite{xiao}, 
and taking $E_{peak}$ = 170 keV and a peak fluance of
$5.7 \times 10^{-4}$ erg cm$^{-2}$~\cite{Mc}, we derive a value for $z$ for GRB041219a of $0.23\pm0.03$. Taking a lower limit of 0.2 for the redshift and
taking $k_2$ = 350 keV/c and $k_1$ = 100 keV/c in Equation~(\ref{constraint}),
we find a new, most accurate cosmological constraint on $|\xi|$ of
\beq
|\xi| \le 2.4 \times 10^{-15},
\label{answer}
\eeq
almost five orders of magnitude better than the previous
best solid limit derived using polarimetric observations of the Crab Nebula in the hard X-ray 
energy range~\cite{Mac}.

From equation \ref{km}, the result given in equation (\ref{answer}) implies a constraint on the isotropic dimension 5 SME parameter of
\beq 
k^{(5)}_{(V)00} \le 4.2 \times 10^{-34}\ {\rm GeV}^{-1}.
\eeq
Finally, it should be noted that with the redshift dependence
obtained from Equations (\ref{integral}) and (\ref{final}), any reasonable redshift for a GRB similar to GRB041219a and showing detectable polarization will give a constraint on $|\xi|$ below $\sim 5 \times 10^{-15}$ corresponding to 
a constraint on $k^{(5)}_{(V)00}$ below $\sim 10^{-33}$ GeV$^{-1}$. 
This can be seen from Figure~\ref{bigD}.

Much better 
tests of birefringence can be performed by polarization
measurements at higher $\gamma$-ray energies. The technology
for measuring polarization in the 5 to 100 MeV energy range
using gas filled detectors is now being developed and tested
~\cite{SH}. Studies of cosmological
sources such as a GRBs at such energies can probe values of
$|\xi|$ several orders of magnitude smaller than is presently
possible.

\section{Frame Independent Constraint} 

The vector $n$ in the EFT model given by equation (\ref{deltaL})
leads to strictly isotropic physics only in one special frame,
usually taken to be the frame in which the cosmic microwave background is isotropic.
In other frames the dispersion relation will have anisotropic components. This
can be taken into account by using the general SME formalism~\cite{KM09}. There are then 16 
independent $k^{(5)}_{(jm)}$ parameters that are weighted by spherical harmonic
coefficients according to their spin weight with respect to the line of sight 
unit vector ${\bf n}$. For GRB041219a this leads to the frame-independent constraint

\begin{equation}
|\sum_{jm}{Y_{jm}}(37^{\circ},0^{\circ})k^{(5)}_{(V)jm}| \le 1.2 \times 10^{-34}\ {\rm GeV^{-1}}.
\end{equation}

\section {Other constraints and Implications}

The Lorentz violating dispersion relation
(\ref{disp-ph}) implies that the group velocity of photons,
$v_{g}=1\pm\xi p/M_{Pl}$, is energy dependent. This leads to an energy
dependent dispersion in the arrival time at Earth for photons spread 
over a finite energy range originating in a distant source. 
The result obtained from observations of the $\gamma$-ray energy-time
profile by the {\it Fermi} satellite for the burst GRB090510 gives 
a limit of $\xi < 0.82$~\cite{rel}. Thus, the time of
flight constraint from {\it Fermi}, while still significant because
it gives $\xi < 1$, remains many orders of magnitude weaker than the
birefringence constraint. However, the {\it Fermi} constraint is
independent of the EFT assumption of helicity dependence of the
group velocity.
Perhaps the best constraint on LIV in general comes from a
study of the highest energy cosmic rays, giving a limit of
$4.5 \times 10^{-23}$ in the hadronic sector~\cite{st09}. 

Thus, all of the present astrophysical data point to the conclusion
that LIV does not occur at the level $\xi(E/M_{Pl})$ with $\xi$ = $\cal{O}$(1). 
In fact, in appears that $\xi \ll 1$. What this is telling us about 
the natures of space-time and gravity at the Planck scale is still an open question.


\section*{Acknowledgements}
I would like to thank Neil Gehrels, Stanley Hunter, Sean Scully, Takanori Sakamoto, Tonia
Venters, and an anonymous referee for helpful discussions.
\newpage


\end{document}